\documentclass[aps,preprint]{revtex4}%
\usepackage{amsfonts}
\usepackage{amsmath}
\usepackage{amssymb}
\usepackage{graphicx}%
\begin{document}
\title[Aharonov-Bohm phase shift]{Unresolved Classical Electromagnetic Aspects of\\the Aharonov-Bohm Phase Shift}
\author{Timothy H. Boyer}
\affiliation{Department of Physics, City College of the City University of New York, New
York, New York 10031}

\begin{abstract}
The long-standing controversy regarding the Aharonov-Bohm phase shift is
reviewed. \ The shifts of both optical and particle interference patterns are
summarized. \ It is pointed out that a line of electric dipoles and a line of
magnetic dipoles (a long solenoid) both produce experimentally observed phase
shifts similar to that produced by introducing a rectangular block of glass
behind one slit of a double-slit interference pattern; the double-slit pattern
is shifted while the single-slit envelope remains undisplaced. \ The quantum
explanation for the magnetic interference pattern shift introduced by Aharonov
and Bohm in 1959 involves completely different ideas from those suggested by a
semiclassical analysis. \ Experiments planned by Caprez, Barwick, and Batelaan
should clarify the connections between classical and quantum theories in
connection with the Aharonov-Bohm phase shift. \ 

\end{abstract}
\maketitle

\subsection{Introduction}

The foundations of quantum theory are intimately connected with our
understanding of the connections between classical and quantum physics. \ One
of the departures between classical and quantum theories occurred in 1959 when
Aharonov and Bohm proposed a new phase shift occurring when charged particles
pass around opposite sides of a long solenoid.\cite{AB} \ The experimentally
observed Aharonov-Bohm phase shift is said to occur in the absence of
classical electromagnetic forces and is said to represent a quantum
topological effect which has no classical analogue. \ 

Although the Aharonov-Bohm phase shift is well-verified
experimentally,\cite{C}\cite{MB} it has been suggested repeatedly during the
past thirty-five years that the phase shift may arise not due to "quantum
topology" but rather may be due to classical electromagnetic interactions
between the passing charges and the solenoid.\cite{B73b}\cite{B87}%
\cite{AC}\cite{B00a}\cite{B00b}\cite{B02a}\cite{B02b}\cite{B06a}\cite{B06b}
\ It is this view which has been sharpened steadily and which very recently
has come under experimental investigation.\cite{Batep} \ Here we wish to
discuss the shifts in particle interference patterns which can arise from
electric and magnetic fields.

\subsection{Interference Pattern Shifts}

Particle inference patterns formed when monoenergetic particles pass through
two slits to a distant screen are analogous to those obtained when
monochromatic light passes through two slits in a wall to a distant screen.
\ In both cases we find a single-slit envelope (associated with the
interference of waves passing through a single slit), and then inside the
single-slit envelope there is a double-slit interference pattern (associated
with the interference of waves passing through two different slits). \ The
interference patterns can be shifted in two distinct arrangements: i)
deflection of the entire interference pattern, both the single-slit envelope
and the double-slit pattern, or ii) shift of the double-slit pattern leaving
the single-slit envelope undisplaced.

i)If a single large prism is place directly behind the two slits so that all
light through the slits also passes through the prism, then the interference
pattern as a whole is deflected, both the single-slit envelope and the
double-slit pattern. \ An analogous deflection of the particle interference
pattern arising from charged particles can be obtained by placing a uniform
electric field parallel to both the\ wall with the slits and to the plane
formed by the particle paths through the two slits, or else by placing a
uniform magnetic field perpendicular to the plane formed by the the charged
particle paths through the slits. \ 

ii)If a rectangular piece of glass is placed behind\ only one of the slits,
then the optical interference pattern will suffer a shift in the double-slit
interference pattern leaving the single-slit envelope undisplaced. \ Indeed,
if the piece of glass is made sufficiently thick or has a sufficiently large
index of refraction so that the waves passing through the glass suffer a
spatial delay larger than the coherence length of the light, then the double
slit interference pattern will break down while the single slit envelope will
remain unchanged. \ Just such a shift (of the double-slit interference pattern
leaving the single-slit envelope undisplaced) can be obtained by placing a
line of electric dipoles or a line of magnetic dipoles (a long solenoid)
between the charged particle beams passing through the two slits, the line of
dipoles oriented perpendicular to the plane formed by the two particle
beams.\cite{B06b} \ 

\subsection{Shift Due to a Line of Electric Dipoles}

In the electric case, the dipoles are oriented parallel to the wall containing
the slits and parallel to the plane formed by the particle beams. \ The line
of dipoles can be obtained by using two line charges $\lambda$ separated by a
small distance $2\varepsilon$ giving a dipole moment per unit length
$2\varepsilon\lambda$.\cite{B87} \ Positive charged particles which pass
through one slit are closer to the positive line charge and so are first
repelled while approaching and then attracted back on receding, while the
charged particles which pass through the other slit are closer to the negative
line charge and so are first attracted while approaching and then repelled
while receding. \ Accordingly, the particles through the different slits have
different velocities while passing the line of electric dipoles and so have a
relative displacement when the particles have passed far beyond the line of
dipoles. \ The electrostatic situation is analogous to the optical situation
where (by use of a piece of glass behind one slit) a relative lag is
introduced between the waves which pass through the two slits. \ The electric
energy of interaction of a charged particle $q$ with a line of electric
dipoles is given by
\begin{equation}
U=qV(\mathbf{r}_{q})
\end{equation}
where $V(r_{q})$ is the electrostatic potential of the line of electric
dipoles evaluated at the position $\mathbf{r}_{q}$ of the passing charged
particle. \ The relative lag $\Delta Y~$between the charges passing on
opposite sides of the line of electric dipoles can be calculated either from
the electric fields of the line of electric dipoles acting on the trajectories
of the charged particles, or by balancing the electrostatic interaction
energies against the kinetic energy of the moving charges $mv_{q}\Delta
v_{q}=-U.$ \ The relative spatial lag is given by \cite{B87}%
\begin{equation}
\Delta Y=\frac{4\pi q(2\varepsilon\lambda)}{mv_{q}^{2}}%
\end{equation}
where $(2\varepsilon\lambda)$ represents the electric dipole moment per unit
length. \ Associated with this spatial lag is also a relative time lag%
\begin{equation}
\Delta t=\frac{\Delta Y}{v_{q}}=\frac{4\pi q(2\varepsilon\lambda)}{mv_{q}^{3}}%
\end{equation}
The relative phase shift follows semiclassically as%
\begin{equation}
\Delta\phi=\frac{mv_{q}\Delta Y}{\hbar}=\frac{4\pi q(2\varepsilon\lambda
)}{\hbar v_{q}}%
\end{equation}
Exactly this same phase shift can be obtained by inserting the electrostatic
potential energy expression $qV(\mathbf{r}_{q})$ for a line of electric
dipoles into the Schroedinger equation and solving by the WKB approximation
assuming that the electrostatic interaction energy is small compared to the
kinetic energy of the particles.\cite{B02b} \ This phase shift was observed
experimentally by Matteucci and Pozzi.\cite{MP}

\subsection{Shift Due to a Line of Magnetic Dipoles (a Solenoid)}

In the magnetic case, a long solenoid is placed between the charged particle
beams which pass through the slits so that the axis of the solenoid (and so
too the magnetic dipoles) are perpendicular to the plane formed by the
particle beams. The magnetic energy of interaction of a charged particle $q$
passing a solenoid with constant currents is given by\cite{B1973}
\begin{equation}
U=\frac{q}{c}\mathbf{v}_{q}\cdot\mathbf{A(r}_{q})
\end{equation}
where $\mathbf{v}_{q}$ is the velocity of the passing charge and
$\mathbf{A}\left(  \mathbf{r}_{q}\right)  $ is the vector potential in the
Coulomb gauge of the solenoid evaluated at the position of the passing charge.
\ We can calculate a possible relative lag between the charged particles which
pass on opposite sides of the solenoid by using the net Lorentz forces which
the passing charges place on the solenoid and then assuming Newton's third law
for the forces on the particles, or else by assuming that energy conservation
for the magnetic energies occurs by changes in the kinetic energies of the
passing particles $mv_{q}\Delta v_{q}=-U.$ \ The possilbe relative spatial lag
is given by\cite{B87}%
\begin{equation}
\Delta Y=\frac{q\Phi}{mv_{q}c}%
\end{equation}
for particles passing on opposite sides of the long solenoid of flux $\Phi$
and magnetic moment per unit length $\Phi/(4\pi).$ \ Associated with this
spatial lag is also a relative time lag%
\begin{equation}
\Delta t=\frac{\Delta Y}{v_{q}}=\frac{q\Phi}{mv_{q}^{2}c}%
\end{equation}
The relative phase shift follows semiclassically as%
\begin{equation}
\Delta\phi=\frac{mv_{q}\Delta Y}{\hbar}=\frac{q\Phi}{c\hbar}%
\end{equation}
Alternatively, one can obtain this same phase shift by introducing the vector
potential $\mathbf{A(r}_{q})$ of the solenoid into the Schroedinger equation
and solving using the WKB approximation assuming that the kinetic energy of
the particles is large compared to the magnetic interaction energy
$(q/c)\mathbf{v}_{q}\cdot\mathbf{A(r}_{q}).$\cite{B02b} \ This phase shift in
Eq. (8) was predicted by Aharonov and Bohm\cite{AB} and first observed
experimentally by Chambers.\cite{C}

\subsection{Comparison of the Electric and Magnetic Phase Shifts}

We have emphasized that the phase shifts due to a line of electric dipoles and
to a line of magnetic dipoles (a solenoid) have both been observed
experimentally, and both take the same form corresponding to a shift in the
double-slit interference pattern while leaving the single-slit envelope
undisplaced. \ Thus these shifts look just like the optical shift obtained by
placing a rectangular piece of glass behind only one slit of the two-slit
interference arrangement. \ Furthermore, semiclassical arguments involving
energy conservation lead to exactly the observed phase shifts. \ This suggests
that both phase shifts may involve classical lag effects arising from
classical electromagnetic forces and velocity changes for the passing charged
particles. \ 

However, this is not at all the accepted explanation in the physics
literature. \ The electrostatic situation (in those rare instances when it is
acknowledge in the literature) is indeed accepted as involving classical
electric forces and lags analogous to the optical case. \ However, it is the
magnetic case\ (the Aharonov-Bohm phase shift) which receives virtually all
the attention in the literature.\cite{Sak} \ It receives this attention
precisely because it is supposed to represent a departure from familiar
connections with classical physics. \ The magnetic situation is said to
involve a quantum topological effect having no classical analogue. \ The
magnetic phase shift is said to occur in the absence of classical
electromagnetic forces on the passing charged particles, and therefore is said
to occur without any velocity changes or relative lags for the passing
charges. \ The novelty of the accepted point of view is emphasized by
Silverman who regards the Aharonov-Bohm phase shift as one of the new
"mysteries" of quantum theory.\cite{Sil} Indeed, the quantum topological point
of view is so deeply entrenched that the lag point of view\cite{B73b} was
published only with difficulty in the Physical Review in 1973; in more recent
years, the Physical Review has rejected all manuscripts giving further
information about the magnetic phase shift from a classical perspective.

\subsection{Experimental Implications}

The classical and quantum analyses for the magnetic phase shift involve
totally different perspectives and lead to different experimental
implications. \ The velocity changes, longitudinal displacements, and time
lags implied by the classical analysis are all different from the quantum
topological point of view which involves no forces or velocity changes. \ Thus
all these aspects become available for experimental discrimination between the
two analyses. \ There is also a further distinction between the classical and
quantum analyses. \ According to presently accepted quantum theory, the
Aharonov-Bohm phase shift arises from an enclosed magnetic flux with no need
to discuss any interaction between the passing charges and the sources of the
magnetic flux. \ In contrast, the classical lag analysis depends crucially
upon the details of the interaction between the passing charged particles and
the current-carriers of the magnetic flux.\cite{B06a} \ The quantum point of
view leads to the expectation that the interaction of a passing charge and a
solenoid should remain the same no matter what the size of the solenoid and no
matter how the solenoid currents are produced. \ Indeed, this perspective
encourages experimentalists to search for a classical lag effect for charged
particles passing a macroscopic solenoid. \ Such an experiment was performed
recently by Caprez, Barwick, and Batelaan, and no classical lag was
seen.\cite{Bate} \ However, from the classical perspective involving the
detailed interaction of charges, large and small solenoids may not behave the
same way; resistance in the wire of the solenoid can make a difference as to
whether or not one should expect to observe an energy-conserving force back on
the passing charges.

One may ask how two such totally different points of view can lead to the same
experimental prediction for the Aharonov-Bohm phase shift. \ The relative
success of the contrasting points of view may reflect a situation analogous to
that when a charged particle passes a conductor. \ As long as the conductor is
a good conductor, the forces on a passing charged particle due to image
charges will involve only the physical shape of the conductor and not the
detailed composition of the conductor. \ However, if the conductivity is
imperfect or if the induced charges correspond to dielectric behavior, then
more information regarding the interaction of the passing charge and the
material will be required to account for the behavior of a passing charged
particle. \ In a similar vein for the Aharonov-Bohm phase shift, it may well
be that when the resistive energy loss of a solenoid is negligible, then all
solenoids behave in the same way regarding energy conservation for a passing
charged particle. \ However, when the resistive energy loss is large, then the
energy-conserving interaction becomes small, and the interaction of the
particle and solenoid becomes quite different. \ If the frictional forces
(solenoid resistance) are large, then we do not expect to find the time lag
because the accelerations of the solenoid charges will be small and hence the
back forces at the passing charge will be small.

In the experiments of Moellenstedt and Bayh,\cite{MB} where the Aharonov-Bohm
phase shift is clearly present, the passage times of electrons past the
solenoids is of the order of $10^{-13}$ sec (for a 40 keV electron passing 20
microns from the center of the solenoid). This time is not much longer than
the collision time $10^{-14}$ sec in the Drude model for conductivity of a
metal. Indeed, Jackson\cite{Jack312} gives $\gamma$ as of the order of
$10^{13}$ inverse seconds where $-m\gamma v$ is the resistive damping of a
particle of mass $m$ and speed $v$. Thus it is possible that the conservation
of energy involving magnetic fields holds in the short-time regime where
Moellenstedt and Bayh's experiments were performed, yet would not hold for the
much longer passage times for the slower electrons passing the much larger
solenoid in the experiment of Caprez, Barwick, and Batelaan.

\subsection{Crucial Experimental Tests}

The crucial test involving time delay corresponds (in a single experiment) to
observing the Aharonov-Bohm phase shift and yet not observing the time delay
of Eq. (7). Such an observation would rule out the classical lag
interpretation of the Aharonov-Bohm phase shift. For the regimes where the
Aharonov-Bohm phase shift has been observed, the time delays would be
extraordinarily small. Thus for the experiments of Moellenstedt and Bayh, the
time delay in Eq. (7) is of the order of $10^{-21}$ sec.\ 

It is the shifts of particle interference patterns which allow the measurement
of extraordinarily small quantities. \ Thus shift of the interference pattern
by a large flux seems the best way to realistically test for the absence of a
velocity change. \ According to the currently accepted quantum point of view,
the Aharonov-Bohm phase shift is a purely topological shift, and there is no
velocity change and hence no spatial lag for charged particles passing on
opposite sides of a solenoid. \ Accordingly, one can increase the flux in a
solenoid by an arbitrary amount and never break down the particle interference
pattern. \ On the other hand, in the classical lag point of view, a
sufficiently large solenoid flux will cause a relative lag which is larger
than the coherence length associated with the charged particles, and at this
point the observed particle interference pattern should break down, just as it
does in the optical case. \ Batelaan's group is hoping to achieve appropriate
conditions to observe this possible breakdown.\cite{Batep}

In the classical analysis, the observed Aharonov-Bohm phase shift due to a
solenoid (a line of \textit{magnetic} dipoles) is analogous to the observed
Matteucci-Pozzi phase shift\cite{MP} due to a line of \textit{electric}
dipoles.\cite{B02b} \ This is not at all the view of currently accepted
quantum theory, which accepts the idea that the electrostatic phase shift is
due to a classical lag but claims that the magnetic phase shift is due to a
new quantum mechanical effect having no classical analogue. \ Thus ideally one
would like to see both these phase shifts (electric and magnetic) tested at
the same time with the same coherence length for the electron beam. From both
classical and quantum analyses of the electrostatic situation, we believe that
we know exactly what is going on involving electrostatic forces leading to a
relative lag effect producing the Matteucci-Pozzi electrostatic phase shift.
If the interference pattern breaks down for large \textit{electric} dipole
magnitude but not for large \textit{magnetic} dipole magnitude, then (in
contradiction to the classical analysis) the mechanisms for these two
interference pattern shifts are quite different, as is indeed claimed by
currently accepted quantum theory. However, if the interference pattern breaks
down for both the electric and the magnetic phase shifts, the currently
accepted quantum view is in error.


\begin{thebibliography}{99}                                                                                               %


\bibitem {AB}Y. Aharonov and D. Bohm, "Significance of electromagnetic
potentials in quantum theory," Phys. Rev. \textbf{115}, 485-491 (1959).

\bibitem {C}R. G. Chambers, "A shift of an electron interference pattern by
enclosed magnetic flux," Phys. Rev. Lett. \textbf{5}, 3-5 (1960).

\bibitem {MB}G. Moellenstedt and W. Bayh, "Messung der kontinuierlichen
phasenschiebung von Elecktronenwellen in kraftfeldfreien Raum durch das
magnetische Vektorpotential einer Luftspule," Naturwissenschaften \textbf{49},
81-82 (1962).

\bibitem {B73b}T. H. Boyer, "Classical electromagnetic deflections and lag
effects associated with quantum interference pattern shifts: considerations
related to the Aharonov-Bohm effect," Phys. Rev. D \textbf{8}, 1679-1693 (1973).

\bibitem {B87}T. H. Boyer, "The Aharonov-Bohm effect as a classical
electromagnetic lag effect: an electrostatic analogue and possible
experimental test," Il Nuovo Cimento \textbf{100}, 685-701 (1987).

\bibitem {AC}T. H. Boyer, "Proposed Aharonov-Casher effect: Another example of
an Aharonov-Bohm effect arising from a classical lag," Phys. Rev. A
\textbf{36}, 5083-5086 (1987).

\bibitem {B00a}T. H. Boyer, "Does the Aharonov-Bohm effect exist?" Found.
Phys. \textbf{30}, 893-905 (2000).

\bibitem {B00b}T. H. Boyer, "Classical electromagnetism and the Aharonov-Bohm
phase shift," Found. Phys. \textbf{30}, 907-932 (2000).

\bibitem {B02a}T. H. Boyer, "Classical electromagnetic interaction of a point
charge and a magnetic moment: Considerations related to the Aharonov-Bohm
phase shift," Found. Phys. \textbf{32}, 1-36 (2002).

\bibitem {B02b}T. H. Boyer, "Semiclassical explanation of the Matteucci-Pozzi
and Aharonov-Bohm phase shifts," Found. Phys. \textbf{32}, 41-49 (2002).

\bibitem {B06a}T. H. Boyer, "Darwin-Lagrangian analysis for the interaction of
a point charge and a magnet: Considerations related to the controversy
regarding the Aharonov-Bohm and Aharonov-Casher phase shifts," J. Phys.
A:Math. Gen. \textbf{39}, 3455-3477 (2006)

\bibitem {B06b}T. H. Boyer, "Proposed experimental test for the paradoxical
forces associated with the Aharonov-Bohm phase shift," Found. Phys. Lett.
\textbf{19}, 491-498 (2006).

\bibitem {Batep}G. Gronniger, Z. Simmons, S. Gilbert, A. Caprez, and H.
Batelaan, "The Aharonov-Bohm effect, phase or force," (preprint).

\bibitem {MP}G. Matteucci and G. Pozzi, "New diffraction experiment on the
electrostatic Aharonov-Bohm effect," Phys. Rev. Lett. \textbf{54}, 2469-2470 (1985).

\bibitem {B1973}T. H. Boyer, "Classical electromagnetic interaction of a
charged particle with a constant-current solenoid," Phys. Rev. D \textbf{8},
1667-1697 (1973).

\bibitem {Sak}See for example, J. J. Sakurai, \textit{Modern Quantum Mechanics
Revised Edition} (Addison-Wesley, Reading, MA 1994).

\bibitem {Sil}M. P. Silverman, \textit{More Than One Mystery: Explorations in
Quantum Interference} (Springer, New York 1995).

\bibitem {Bate}A. Caprez, B. Barwick, and H. Batelaan, "A macroscopic test of
the Aharonov-Bohm effect," (preprint).

\bibitem {Jack312}J. D. Jackson, \textit{Classical Electrodynamics 3rd ed}
(Wiley, New York 1999), p. 312.
\end{thebibliography}
\end{document}